\documentclass[pre]{revtex4}
\usepackage[english]{babel}
\usepackage{amsmath}
\usepackage{epsfig}

\begin{document}

\title{\bf FROM COULOMB TO THE EFFECTIVE INTERACTION: APPLICATION TO BOSE CONDENSATION
}
\author{S.A. Trigger $^{1,2}$}
\address{$^1$ Joint\, Institute\, for\, High\, Temperatures, Russian\, Academy\,
of\, Sciences, Izhorskaia St., 13, Bd. 2. Moscow\, 125412, Russia;\\
$^2$ Institut f\"ur Physik, Humboldt-Universit\"at zu Berlin,
Newtonstra{\ss}e 15, D-12489 Berlin, Germany.\\
e-mail: satron@mail.ru
}

\begin{abstract}

The expression for the short-range effective interaction potential of "quasinulei"\, is derived based on the model of the "pure"\, Coulomb interaction. This model represents the equilibrium Coulomb system (CS) of interacting electrons and the identical nuclei, using the adiabatic approximation for nuclei and an arbitrary strong (in general) interaction for the electronic subsystem. On the basis of general properties of Coulomb interaction it is shown that the Fourier-component of the pair effective potential between "quasinuclei" possesses discontinuity at the wave vector $q=0$. This discontinuity is essential for the Bose condensed systems as HeII and the rarified Alkali metals at temperatures lower than the Bose condensation transition, since there are macroscopic quantity of quasiparticles with the momentum $q=0$. In particular, it is shown that for the single-particle excitations can exist the gap which disappears in the normal state. The value of this gap is estimated.
The problem of generalization of the obtained results for the case of a strong electron-nuclei (or electron-point ion system) is discussed. \\

PACS number(s): 51.30.+i, \emph{52.25.Kn}, 31.15.-p\\

\end{abstract}

\maketitle

\section{Introduction}

Exact models for describing equilibrium systems of interacting particles are of fundamental importance.
The most adequate model of the real matter is the quasi-neutral non-relativistic system of interacting electrons and nuclei. Hence, the matter properties are almost completely controlled by the Coulomb interaction collective behavior of electrons and nuclei interacting via Coulomb forces. This approach in statistical physics appeared in the pioneering papers [1-4]. The modern view on the problem and the extended list of references described in [5].

Consideration of matter in the framework of CS is the most appropriate since the fitting parameters of the interaction potential are absent.
However, even in the simplest case of identical nuclei the existing of only four independent parameters, namely the interaction parameter $\Gamma=e^2 n_e^{1/3}/T$, the degeneration parameters for electrons and nuclei $\lambda_e=\hbar^2 n_e^{2/3}/m_e T$, $\lambda_i=\hbar^2 Z n_e^{2/3}/m_i T$ and the nuclei charge $Z$ leads to huge diversity of possible states and peculiarities for the properties of CS, which cannot be considered without specific approximations. For hydrogen, when $Z=1$ the situation is more simple and can be represented by a convenient diagram introduced in [6] (see also, e.g.,  [7]). Later on this diagram has been extended for consideration of various more complicated CS [8], including quark-gluon plasma (see, e.g., [5]). The very recent discussion on the perturbation row for thermodynamics of CS is presented in [9,10] (see also [11]).

In many cases, at relatively low temperatures and densities, the model of the "simple"\, (or "neutral"\,, or "ordinary") matter is used, which is a system of identical composite "particles"\, (in fact quasiparticles for the statistical systems as, e.g., gas of atoms), interacting with each other via the effective short-range potentials.
The model of the "simple"\, matter can be considered as a useful approach based on the introduction of quasiparticles (e.g., "atoms"\, in gases and condensed phases of matter) appropriate for a certain range of thermodynamic parameters [1,12]. The statistical theory of the "simple"\, matter is well developed on under the assumption that interaction potentials of "atoms"\, are known a priori (see, e.g., [12, 13]). This leads to significant ambiguity in the determination of interaction potentials of such quasiparticles, e.g., in the case of consideration of the "atom"\, - "atom"\, interaction, when there is the necessity to use the fitting parameters appears.

The determination of the effective interaction potential of "atoms"\, is directly associated with the problem of development of the "simple"\, matter model. Transition from the concept of the "pure"\, matter (or the "physical"\, model of matter, representing a quasi-neutral two-component system with the Coulomb interaction, consisting of electrons and nuclei of the same type) to the "chemical"\, model of substance (which introduces atoms and molecules as the basis components)  is a complicated problem.

So far, the interaction potentials of "atoms"\, in a many-body system of nuclei and electrons are determined (see [14,15] and references therein) either

(i) within the quantum-mechanical problem of determining the interaction potential of two nuclei, taking into account electrons localized at them, whose number is equal to the total charge of nuclei under consideration. Here we have to stress that in this case we fully disregard all interactions between the atoms which are more complex than the pair ones, and may lose control on collective effects or\\
(ii) within the problem of electronic states in Wigner--Seitz-type "cells"\, when considering solids
or\\
(iii) within the Bloch delocalized waves the interaction between electrons and the immobile lattice is taken into account precisely, but we loose the flexibility and cannot discuss anything except for ideal crystal.\\
As is known, for a large class of CS the condition of quasineutrality $n_e=Z_n n_i$ (where $Z_n$ is the charge number for nucleus or the point ion) which provides the elimination (nullification) of the divergent terms $\sum_{a,b}n_a n_b v_{a,b}(q=0)$ in the row of the perturbation theory is fulfilled. Here the indexes $a,b$ are the notation for electrons and nuclei interacting via the Coulomb potentials $v_{a,b}(q)=e_a e_b /q^2$ (or for electrons and the point ions of charge $Z$ in an electron-ion plasma). However, in [16] it was shown that to provide the matching of results for the canonical ensemble and the Grand ensemble the quasineutrality condition is not enough. The more strong condition for Coulomb potentials
\begin{eqnarray}
v_{a,b}(\textbf{q}=0)=0
\label{F1}
\end{eqnarray}
is necessary to construct the self-consistent description of CS.
It seems that condition (1) was first used in Ref.[2] when considering thermodynamic properties of weakly nonideal plasma within the Grand ensemble.

This condition has a deep physical sense.
 .
The fact is that the establishment of the explicit form of the interaction potential between charged particles (Coulomb's law) as a function of distance between them in field theory is based on the Fourier transform of Maxwell's equations [17]. According to the field theory, the Coulomb potential determines the electrostatic interaction of charged particles, therefore, to determine the Fourier transform of the potential of the Coulomb interaction potential, the Poisson equation is used, which leads to the following result for $q\neq 0$
\begin{eqnarray}
v_{a,b}(\textbf{q})=\frac{4\pi e_a e_b}{q^2}
\label{F2}
\end{eqnarray}
In this case, the value of $v_{a,b}(\textbf{q}=0)$ remains uncertain, according to the Poisson equation.
In such uncertainty, at first glance, there is no physical reason, since according to classical field theory, the electric field strength, which is determined by non-zero wave vectors, has a direct meaning [17].
In addition, in view of (2), it is easy to establish the Coulomb law using the integral Fourier transform

\begin{eqnarray}
v_{a,b}(\textbf{r})=\int \frac{d^3 q}{(2\pi)^3} \exp (i \textbf{q} \textbf{r})v_{a,b}(\textbf{q})
\label{F3}
\end{eqnarray}
We notice, that in this case condition (1) is no way affects the form of the Coulomb interaction potential $v_{a,b}(r)$ in the
$\textbf{r}$-space.

However, as it is shown in [16] (see also [18], [19]) equality (1) cannot be rigorously justified in the framework of classical theory and requires consideration on the basis of quantum electrodynamics. Thus, when
constructing the statistical theory for the quantum non-relativistic CS, instead of the integral Fourier
transform (2) for the Coulomb potential, the Fourier series
\begin{eqnarray}
v_{a,b}(\textbf{r})= \frac{1}{V} \sum_q \exp (i \textbf{q} \textbf{r})\, v_{a,b}(\textbf{q})=\frac{e_a e_b}{r}
\label{F4}
\end{eqnarray}
should be used.
To date, in calculating the value of $v_{a,b}(\textbf{q}=0)$, in most cases, it is conventional to proceed from the
fact that the potential $v_{a,b}(\textbf{r})$ is known in the sense that its value is defined by the experimentally
determined Coulomb law. In this case, according to the definition of the Fourier transform for the
potential $v_{a,b}(\textbf{r})$ (3)
\begin{eqnarray}
v_{a,b}(\textbf{q}=0)=\int_V d^3 r v_{a,b}(\textbf{r}), \qquad \lim_{V\rightarrow \infty}V^{-1}v_{a,b}(\textbf{q}=0)=0,
\label{F5}
\end{eqnarray}
since the integral in (4) diverges as $V^{2/3}$.
On this basis, when determining thermodynamic properties of weakly nonideal plasma in [2],
the statement $v_{a,b}(\textbf{q}=0)=0$ was formulated from intuitive considerations, taking into account the quasi-neutrality condition $\sum_{a,b}n_a n_b v_{a,b}(q=0)$ in the thermodynamic limit. A similar statement is also used in the study of the charged Bose gas [20, 21], and
when considering correlation functions and linear electromagnetic properties of the CS [22].

Following to [18],[19], to solve the problem of the value of $v_{a,b}(\textbf{q}=0)=0$, it is necessary to turn to
the results of the quantum field theory (e.g., [23]) according to which charged particles interact with each other
through the quantized electromagnetic field. Within quantum statistical electrodynamics [24], we can
speak of the correspondence between the Green function for the quantized electromagnetic field $D_{\mu,\nu}(k)$
($\mu,\nu = 0,1,2,3$; $k = (\omega/c, \textbf{q})$) and the interaction potentials between charged particles. In so doing,
the discrete momentum representation (see (4)) [11] should be used. In this case, the 4-vector of the
potential corresponding to the quantized electromagnetic field does not contain term with $\textbf{q} = 0$, since
the electromagnetic field quantum energy is $\hbar c|\textbf{q}|$. Therefore, the wave vector $\textbf{q}$ in the Green function
 $D_{\mu,\nu}(k)$ cannot be zero. This means that the transferring momentum $\hbar \textbf{q}$ during the interaction should be
nonzero. In other words, there is no physical substance which is a carrier of the interaction, which leads to
the zero momentum transfer.
In the Coulomb gauge which best corresponds to the transition to the non-relativistic limit in the
description of the systems of charged particles, the Green function $D_{00}^{(0)}(k)$ for the free electromagnetic
field has the form $D_{00}^{(0)}(k)=4\pi/k^2$ which corresponds to the Coulomb interaction potential of charged
particles. In the presence of charged particles, the Green function $D_{00}^{(0)}(k)=4\pi/k^2\varepsilon^l(k)$, where $\varepsilon^l(k)$
is the longitudinal permittivity of the system of charged particles and the quantized electromagnetic
field [23], which is fully consistent with the screened Coulomb interaction of charged particles for the
equilibrium CS in the non-relativistic limit [1]. Thus, according to the results of quantum statistical
electrodynamics, the statement (1) for $v_{a,b}(q = 0)$ is satisfied in the sense that the Coulomb interaction
potential $v_{a,b} (r)$ can be written as the Fourier series (4) in which the term with $q = 0$ is absent. In other
words, the relation $v_{a,b}(\textbf{q }= 0) = 0$ in the CS is valid even without the quasi-neutrality condition. The detailed discussion about
the relations (1) and the quasineutrality condition is recently presented in [5] on the basis of [16], [18].

In connection with the foregoing, the question arises about the influence of relation (1) on the form of effective potential. This question can be essential, in particular, for consideration of the Bose condensed systems, where particles (the composite bosons) with zero momentum plays a crucial role, for the ionized Rydberg plasmas and for other systems.

For the beginning let us consider a simple model of effective potential for the case of a weak Coulomb interaction between electron and nuclei subsystems. Due to the mass difference ($m_e\ll M_n$, where $m_e$ and $M_n$ are the electron and nuclear masses, respectively) we can use the adiabatic approximation.
As well known (see, e.g., [25]) the Fourier transform of the effective potential $v^{eff}_{n,n}$ between ions (nuclei for the "pure"\, CS) in the case under consideration can be written in the form
\begin{eqnarray}
v^{eff}_{n,n}(\textbf{q})= v_{n,n}(\textbf{q})+ v^2_{e,n}(\textbf{q})\frac{\Pi_{e}(\textbf{q})}{\varepsilon^l_e(\textbf{q})}, \qquad \varepsilon^l_e(\textbf{q})\equiv 1- v_{e,e}(q)\Pi_{e}(\textbf{q}),
\label{F6}
\end{eqnarray}
where $\Pi_{e}(\textbf{q})$ is the polarization function for electron liquid with arbitrary strong interaction, $v_{e,n}=-4\pi Z_n e^2/q^2$ and $v_{n,n}=4\pi Z_n^2 e^2/q^2$. According to relation (1) [16] (see also [18],[19] and [5]) the value of $v^{eff}_{n,n}(\textbf{q=0})=0$, since $\Pi_{e}(\textbf{q=0})$ is finite. At the same time for $q\rightarrow 0$ from (6) we arrive at the value $v^{eff}_{n,n}(\textbf{q}\rightarrow 0)$
\begin{eqnarray}
v^{eff}_{n,n}(\textbf{q}\rightarrow 0)= -\frac{Z_n^2}{\Pi_{e}(\textbf{q=0})},
\label{F7}
\end{eqnarray}
Taking into account that the long wavelength limit of the polarization function is related with the screening length $R_{sc}$ by the equality $\Pi_{e}(\textbf{q=0})=-1/(4\pi e_e^2 R^2_{sc})$ we found the discontinuity at the point $q=0$ for the nuclei effective potential, which is equal
\begin{eqnarray}
\Delta = v^{eff}_{n,n}(\textbf{q}\rightarrow 0)-v^{eff}_{n,n}(\textbf{q}\equiv 0)= Z_n^2 4\pi e_e^2 R^2_{sc}.
\label{F8}
\end{eqnarray}
For a weakly interacted degenerate electron subsystem the screening length equals $R_{sc}=1/k_{TF}=(\varepsilon_F/6\pi n_e e^2_e)^{1/2}$, where $\varepsilon_F$ is the Fermi energy. For the typical electron density in metals $n_e\simeq 5\cdot 10^{22}=7,5\cdot 10^{-3}/a^3_0$ the value of $R_{sc}\simeq 1,3\cdot 10^{-8}cm\simeq 2,4 \,a_0$ and the screening is very effective. The characteristic value of $\Delta\simeq Z_n^2\cdot 30,1 \cdot  e_e^2\, a_0^2$ and the characteristic gap energy $E_\Delta$ can be estimated as $E_\Delta \geq \gamma n_n\cdot\Delta \simeq \gamma 0,4 Z_n^3 Ry$, where $\gamma=n_c/n_n$ is the ratio of the quantity of particles (in the case under consideration nuclei) in the
condensate $n_c$ to the full number of particles (nuclei) $n_n$ . This gap between the condensate and excited
quasiparticles provides superfluidity for enough low condensate flow. The appropriate criterion
of superfluidity leads to disappearance of the superfluid motion for $n_c\rightarrow 0$.

According to the above consideration the system under consideration is transferred to the system of "quasinuclei"\, and electrons. The Fourier-component of the effective interaction (interaction between "quasinuclei") possesses a discontinuity at $q=0$. The results of this consideration can be applied to the system of electrons and Bose condensed nuclei [26] and leads to the nuclei superconductivity and Meissner effect for nuclei in this model.

The question arises how to take into account a strong electron-nuclear interaction, when the atoms (or more complicated quasiparticles) appears, constructed from the initial purely Coulomb model of substance? What is the structure of an effective potential for the "atomic quasiparticles"? Is the discontinuity of this potential at $q=0$, which provides the gap in the single-particle spectrum of excitations for Bose gases below the condensation temperature (similar to the case of a weak electron-nuclear interaction considered above)? The way for solution of these problems has been proposed in [27], however, the full consideration still absent and will be presented in a separate paper.

\section*{Acknowledgment}
Author is thankful to V.B. Bobrov, A.M. Ignatov and I.M. Sokolov  for the useful discussions.

\end{document}